\documentclass{arttype}
\usepackage{graphicx}

\usepackage{url}
\usepackage{hyperref} 
\usepackage{amsmath}
\usepackage{graphicx}
\usepackage{rotating}
\usepackage{subfigure}
\usepackage{vectortildebarhat}
\usepackage{qhcDefinitions}
\RequirePackage{epsfig}
\RequirePackage{amssymb}
\RequirePackage{natbib}
\RequirePackage{graphicx}

\begin{document}
\begin{frontmatter}
\title{On a Connection between Entropy, Extensive Measurement and Memoryless Characterization}

\author{Peter Sunehag}
\thanks{The author is grateful to T. Sears and A. Caticha for their constructive comments.}
\address{Statistical Machine Learning Program
\\NICTA
\\Locked bag 8001
\\2601 ACT
\\Australia
\\Peter.Sunehag@NICTA.com.au}

\begin{abstract}
We define an entropy based on a chosen governing probability distribution. If a certain kind of measurements follow such a distribution it also gives us a suitable scale to study it with. This scale will appear as a link function that is applied to the measurements. A link function can also be used to define an alternative structure on a set. We will see that generalized entropies are equivalent to using a different scale for the phenomenon that is studied compared to the scale the measurements arrive on. An extensive measurement scale is here a scale for which measurements fulfill a memoryless property. We conclude that the alternative algebraic structure defined by the link function must be used if we continue to work on the original scale. We derive Tsallis entropy by using a generalized log-logistic governing distribution. Typical applications of Tsallis entropy are related to phenomena with power-law behaviour.
\end{abstract}


\end{frontmatter}


\section{Introduction}

\subsection{Entropies}
There are many different ways to describe entropy. One common way is to view it as a measure of uncertainty. Formally it is usually defined as the expectation of a function that depends only on an event's probability, i.e. we take the expectation of a function $V$ defined on $(0,1]$. In information theory, the function is often described as measuring the information content of the event. For the classical Shannon-Boltzman-Gibbs entropy (SBG entropy) $V(p)=\log{\frac{1}{p}}$. $V$ is then, among other things, non-negative, decreasing, has range $[0,\infty)$ and 
\begin{equation}\label{exten}
V(p_1p_2)=V(p_1)+V(p_2).
\end{equation}
The last property is called extensivity and is a critical assumption that characterizes SBG entropy. In physics, the extensivity property is motivated by the argument that if we have two independent systems, then the total energy of the combined system should be the sum of the energies of the separate systems. In information theory, the function $V$ is thought of as an idealized storage need. An efficient storage system is based on a coding scheme where common events have short descriptions (few bits) and rarer events demand more space. For a fixed coding scheme, the extensivity of the idealized storage need is logical. If we take a finance perspective on $V$, we could view it as the value/cost/price of something, e.g. the cost of a claim (or perhaps the total claims for a quarter) for an insurance company. Again, adding up independent costs is very reasonable.

The extensivity property is an assumption that together with some regularity assumptions leads to SBG entropy, see Jaynes \cite{Jay03} for details. However, during the last decades there has been much interest in other entropies that are not extensive \cite{GelTsa04}. For Tsallis entropy \cite{Tsa88} $S_q$, instead of Eq. \ref{exten} we have the relation $S_q(A,B)=S_q(A)+S_q(B)+(1-q)S_q(A)S_q(B)$ for independent events $A$ and $B$. Tsallis entropy is defined by using a deformed logarithm $\log_q$ and is, therefore, an example of a larger class of entropies that are defined from deformed logarithms. If $\phi:[0, \infty) \rightarrow [0, \infty)$ is a strictly positive and non-decreasing function on $(0, \infty)$, then $\lphi$ is defined by
  \begin{align}
    \label{eq:logphi}
    \lphi(p) : = \int_{1}^{p}  \frac{1}{\phi(y)}~dy.
  \end{align}
  If this integral converges for all finite $ p > 0$, then $\lphi$ is
  a deformed logarithm. 

Tsallis entropy, corresponding to $\phi(p) = p^q$, has been successfully applied to many areas of science and in particular to situations where heavy tailed power-law behaviour is encountered. For a density $f$ it is defined by $S_q(f)=\frac{\int f(x)^q dx-1}{1-q}$.
 Despite the practical successes, there is much debate about the foundations for using non-extensive entropies \cite{Cat07, Nau03, CouSci00, WilWlo00}.    

An entropy is used in modeling to choose one among all the probability distributions that satisfy certain constraints encoding available information. Jaynes, the founder of the principle of maximum entropy as a principle for statistical inference, explained it as choosing the least biased among the distributions, i.e. the distribution that did not include any further assumptions \cite{Jay57}. The entropy itself, however, is introducing an extra assumption since not every distribution can arise through maximizing SBG entropy subject to constraints. The distributions that are compatible with SBG entropy constitute the exponential family. Those distributions have  Cumulative Distribution Function (CDF) of the form
$dF(x|\eta)=e^{-\eta^TT(x)-A(\eta)}dG(x)$
where $G$ is the reference measure, $T=(T_1,...,T_n)$ is a sufficient statistic, $\eta=(\eta_1,...,\eta_n)$ is the natural parameter and $A(\eta)$ is needed for normalization. In the case when the reference measure is the Lebesgue measure on $\mathbb{R}$ and we only use one constraint $E(g(x))=\eta$, we have a family of densities of the form $C_\lambda e^{\lambda g(x)}$. Given a density $f$ with finite SBG entropy we can let $g(x)=\log f(x)$ and thereby define a one parameter family of densities of the form $C_\lambda f(x)^\lambda$. Letting $\eta=\int f(x)\log(f(x))dx$ implies $\lambda=1$. The expectation of $g$ with respect to another density $h$ is the cross entropy between $h$ and $f$.

Although densities are always assumed to be integrable, non-negative and to have integral one, the space where the SBG entropy is finite is a proper subspace of $L_1(\mathbb{R})$. For $q\neq 1$, the Tsallis entropy for a density $f$ is finite if $f\in L_q(\mathbb{R})$. Finiteness of SBG entropy for a density $f$ can be seen as an extra regularity condition which coincides with the condition that is necessary to assure that the density's Hardy-Littlewood maximal function 
$$(Mf)(x)=\sup_{r>0} \frac{1}{\mu(B_r(x))}\int_{B_r(x)}|f(y)|d\mu(y)$$
is a member of $L_1(\mathbb{R})$. $B_r(x)=\{y\ |\ |x-y|<r\}$. For every $q>1$, $M$ is a bounded sublinear operator from $L_q(\mathbb{R}^d)$ to itself. This result of Hardy and Littlewood implies Lebesgue's differentiation Theorem regarding differentiating integrals of integrable functions and Rademacher's Theorem regarding differentiating Lipschitz functions. The maximal operator plays a central role in several areas of modern mathematical analysis including real analysis, harmonic analysis and functional analysis \cite{Guz75}. A functional analysis approach to entropy is to use the characteristic function, $V(p)=\|I_A\|_X$ where $I_A$ is the indicator of $A$ with measure $p$, of a rearrangement invariant (r.i.) normed function space $X$. The r.i. property implies that $\|I_A\|_X$ only depends on $p$.

\subsection{The Approach of This Article}
The purpose of this article is to provide a theoretical approach to defining entropies and probability theory based foundation for using them as well as a method for choosing an entropy that is suitable for a particular modeling task. We reach the conclusion that this is equivalent to finding an appropriate scale for a certain kind of observations.

To understand why a function $V$ that does not satisfy the extensivity assumption \ref{exten} can yet be sensible, we consider the task of modeling dramatic rainfall. Expressions like "10-year storm" often occur when describing such events. If we have $1/10$ probability of having a storm with rainfall amount $V(1/10)$ in any given year then, assuming independence, we would have a $1/100$ probability of having two of those storms. The total rainfall of two such storms would be $2V(1/10)$, which certainly does not have to be the same as the rainfall encountered during a "100-year storm" which is what $V(1/100)$ would denote. This distinction between two independent events with probability $1/10$ and one event with probability $1/100$ is not noticed by the SBG entropy which is extensive. The extensivity that characterizes SBG entropy is related to the memoryless property of the exponential distribution, i.e.
\begin{equation}
Pr(X>a+b)=Pr(X>a)Pr(x>b)
\end{equation}
if and only if the random variable $X$ is exponentially distributed. Note that $V(p)=\log{\frac{1}{p}}$ implies that $V^{-1}(t)=e^{-t}$, which is the survival function (Complementary Cumulative Distribution Function) for the exponential distribution with mean $1$, i.e.
\begin{equation}
V^{-1}(t)=Pr(X>t)
\end{equation}
for a random variable $X$ with such a probability distribution. Thus, $V(p)$ is the answer to the question: At least how large can you say with probability (certainty) $p$ that $X$ will be? This is, in a sense, the information that corresponds to the probability $p$. This question can be answered for any probability distribution on $[0,\infty)$ and we will in this article make that the foundation for defining an entropy.

A key tool to connect a governing distribution to a scale is Ghitany's \cite{Ghi96} generalized memoryless characterization that can be formulated for any probability distribution. Those properties are on the form
\begin{equation}
Pr(h(X)>a+b)=Pr(h(X)>a)Pr(h(X)>b)
\end{equation} 
for a function $h$ that we will call a link function.
We will in this article prove that Tsallis entropy corresponds to the generalized log-logistic distributions that Ghitany paid special attention to.
 
\subsection{Scales}
Using a link function is related to changing the scale. If a measurement results in $X$, we are instead really studying $h(X)$ or if we are using a statistic $T$ and are estimating a vector $\eta$ of parameters, we would study $h(\eta^TT(x))$ instead of $n^TT(x)$.
 
One famous example of a scale is the Richter scale, which is used to measure the strength of an earth quake. Its a base-10 logarithmic scale calculated from displacements from zero on a seismometer output.

When it comes to our rainfall example, the distribution should clearly not be memoryless in the ordinary sense. If we have had $0$ mm of rain by noon on a specific day the probability that we would get $x$ mm after noon is smaller than the probability that we would get an additional $x$ mm of rain that day if we have had $a>0$ mm already. If we change the scale in the correct way, we would, however, be able to be memoryless on that scale.

In the theory of Extensive Measurement \cite{KraLucSup71,RobLuc68}, scales are functions $h:G\to\mathbb{R}_+$ on ordered semigroups with the properties 1. $h(x)<h(y)$ iff $x<y$ and 2. $h(x\times y)=h(x)+h(y)$ if $\times$ is the group operation and $+$ is the usual addition in $\mathbb{R}_+$. In this article, we will use a chosen $h$ to define a group operation which makes $h$ additive instead of the other way around.

Albert Einstein rescaled measurements of the fundamental extensive quantities, e.g. he defined the distance between two points based on how long time it would take light to travel from one to the other. In the presence of curved space, this do not coincide with the euclidean distance between the points.

\subsection{Outline}
In chapter two we review various approaches to describing the behaviour that governs a system and in chapter three we review Ghitany's generalized memoryless property. Chapter four deals with analogue algebra like $q$-deformations and it describes how a link function can be used to either define a memoryless property of a distribution or to deform the elementary operations and chapter five is about Tsallis entropy. Chapter six summarizes the general results in the form of theorems and chapter seven concludes our discussion of rainfall modeling that is used as an example throughout the article. Chapter eight contains some concluding remarks and chapter nine is a short summary.

\section{Governing a System}
 
\subsection{Governing Distributions and Superstatistics}
The idea to use a probability distribution to define an entropy has been proposed recently by Niven \cite{Niv07}. Niven used a combinatorial approach where he returns to the origins of SBG entropy. Consider a gas of $n$ particles in an enclosed space that we have divided into $N$ cells of equal size. Using the assumption that the particles will distribute themselves over the cells according to a uniform multinomial distribution is the beginning of the argument that leads to SBG entropy. Niven considered various other alternatives including Polya urn models.

A Polya urn can be described by considering an urn containing balls of various colours, e.g. one ball each of $N$ different colours. When a ball is drawn from the urn, we note its colour, we put it back and add an extra ball of that colour. Thereby we increase the probability of drawing that colour the next time. The expected proportions after the next draw always equal the current proportions, i.e. the process satisfies the Martingale property. The colours here represent different cells and the balls represent particles. If we let the number of drawn balls $n$ tend to infinity, we  end up with a multinomial since the added ball becomes a smaller and smaller proportion of the total number of balls. Which multinomial we will end up with is, however, not predetermined. The resulting proportions follow a Dirichlet distribution and the Bayesian formulation of a Polya urn is to have a Dirichlet prior on the family of multinomials. Therefore, it is sometimes called the Dirichlet Compound Multinomial. The parameters are the initial number of balls in the urn, e.g. $\alpha>0$ of each colour. Polya urns have the important property that they are infinitely exchangeable, i.e. the probability of a finite sequence of any length does not depend on the order. 

The Dirichlet distributions are the conjugate priors to the multinomials. A Dirichlet distribution can also be described as forming a vector of proportions by dividing each of $N$ independent identically Gamma distributed random variables by their sum, i.e. normalizing the vector consisting of the original variables. 

This observation leads us to the Beck-Cohen superstatistics \cite{BecCoh03} idea of defining a distribution for one cell. They pick a probability density function $f(\beta)$ on $(0,\infty)$ where $\beta$ is the inverse temperature, and then normalize by a generalized Boltzmann factor defined by 
\begin{equation}\label{Bfact}
B(E)=\int_0^\infty f(\beta)e^{-\beta E}d\beta.
\end{equation}

Tsallis statistics has been derived as a special case of superstatistics by letting $f$ be the density of a Gamma distribution. Superstatistics has been successfully applied e.g. to model turbulent systems with temperature fluctuations. In superstatistics, a cell has a distribution over temperatures instead of a fixed temperature and can, therefore, incorporate fluctuating local temperatures. The governing distribution approach used in this article is not identical to the superstatistical one, since we work on the other end of the Laplace transform \ref{Bfact}. $B(E)$ is the Laplace transform of the density $f(\beta)$ and from that it follows that $B(0)=1$, that $B(E)$ is decreasing and tending to $0$ as $E$ tends to infinity. $B(E)$ is, therefore, a survival function and we can find $E$ as a function of $p=B(E)$. This recovers our approach. We have a generalized log-logistic distribution as our governing distribution for Tsallis statistics.

Note that Equation \ref{Bfact} employs the Maxwell-Boltzmann distribution, which assigns probabilities for the kinetic energy of a particle as a function of the temperature. Therefore, defining a  probability distribution over temperatures is to define probabilities for probabilities of energy which is Bayesian statistics. Superstatistics is just a particular way of expressing and understanding the prior that is suitable for thermodynamics.

\subsection{Governing Priors}
Instead of beginning with an assumption that particles are distributing themselves over cells according to a uniform multinomial or another predetermined multinomial, we can assume that they are distributed according to a multinomial that is drawn from a Dirichlet distribution. We can describe this as having a distribution that is a mixture of an infinite number of multinomials. Sampling from such a distribution can be performed using the Polya urn scheme described in the previous section. In that section we mentioned that Polya urn models are infinitely exchangeable or in other words, "Bag of Particle Models". According to the De Finetti Theorem, every infinitely exchangeable distribution can be represented as a mixture of multinomials, i.e. with a prior.

\subsection{Governing Dynamics}
As discussed recently by Cohen \cite{Coh05}, there is a connection between dynamics and statistical mechanics. The dynamics describe a particle theory, which is used to define the statistics. SBG entropy is based on the differential equation
\begin{equation}
\frac{dy}{dx}=y,\ y(0)=1
\end{equation}
whose solution $y(x)=e^x$ can also be written as $x=\log(y)$. To define a governing distribution we let
\begin{equation}
Pr(X>x)=\frac{1}{y(x)}.
\end{equation}
If we instead start from
\begin{equation}
\frac{dy}{dx}=\phi(y),\ y(0)=1
\end{equation}
with $\phi$ as in \ref{eq:logphi}, the result is
\begin{equation}
x(y) = \int_{1}^{y}  \frac{1}{\phi(s)}~ds
\end{equation}
and letting $p=\frac{1}{y}$ and $V(p)=x(y)$ implies that
\begin{equation}
V(p)=\lphi(\frac{1}{p}).
\end{equation}
If we refer to $y=\frac{1}{p}$ as the surprise and $x(y)$ as the cost of the
surprise, e.g. in storage demand with the coding scheme that is being used, then the differential equation governs how the cost increases with the surprise.

\section{Characterization Of The Generalized Log-Logistic Distribution}\label{genloglog}
This section is based closely on Ghitany's presentation \cite{Ghi96}.
The exponential distributions have Cumulative Distribution Function (CDF)
\begin{equation}
F(t)=1-e^{-\frac{t}{\lambda}},\ \lambda>0,\ t>0
\end{equation}
and, therefore, survival function
\begin{equation}
\bar{F}(t)=1-F(t)=e^{-\frac{t}{\lambda}}.
\end{equation}
An alternative way of characterizing exponentially distributed random variables is by the memoryless property
\begin{equation}
Pr(X>a+b)=Pr(X>a)Pr(X>b).
\end{equation}
Other distributions also have similar characterizing properties. If $X$ has CDF $F$, then both $F(x)$ and $\bar{F}(X)$ are uniformly distributed on the interval $(0,1)$. Therefore, both $G(x)=\log((F(X))^{-1})$ and $\bar{G}(X)=\log((\bar{F}(X))^{-1})$ are exponentially distributed with $\lambda=1$. Thus, as Ghitany shows, $F$ has the characterizing property
\begin{equation}\label{prop1}
Pr(\log((F(X))^{-1})>a+b)=Pr(\log((F(X))^{-1})>a)Pr(\log((F(X))^{-1})>b)
\end{equation}
or we can use $\bar{F}$ instead of $F$. If $X$ satisfies \ref{prop1} there
is $\lambda>0$ such that $Pr(G(X)>t)=e^{-\lambda t}$.
Ghitany considered two classes of generalized log-logistic distributions, namely $GLLD_1$ defined by
\begin{equation}
F(t)=\frac{1}{(1+(\frac{t}{\alpha})^{-\beta})^m},\ t>0,\ \alpha,\beta,m>0
\end{equation}
and  $GLLD_2$ defined by
\begin{equation}
F(t)=1-\frac{1}{(1+(\frac{t}{\alpha})^{\beta})^n},\ t>0,\ \alpha,\beta,n>0.
\end{equation}
The $GLLD_1$ characterizing property becomes $Pr(\log((1+(\frac{X}{\alpha})^{-\beta})^m)>a+b)=$
\begin{equation}\label{glld1prop}
Pr(\log(1+(\frac{X}{\alpha})^{-\beta})^{m})>a)Pr(\log(1+(\frac{X}{\alpha})^{-\beta})^{m})>b)
\end{equation}
and $GLLD_2$ is, using $\bar{F}$, characterized by $Pr(\log((1+(\frac{X}{\alpha})^{\beta})^n)>a+b)=$
\begin{equation}\label{glld2prop}
Pr(\log(1+(\frac{X}{\alpha})^{\beta})^{n})>a)Pr(\log(1+(\frac{X}{\alpha})^{\beta})^{n})>b).
\end{equation}
The properties do not depend on $n$ and $m$ and, therefore, we can e.g. choose $m=n=1$. or $\frac{1}{m}=n=\alpha$. Using $m=n=1$, Ghitany provided the elegant characterization
\begin{equation}
Pr(1+(T/\alpha)^{-\beta}>xy)=Pr(1+(T/\alpha)^{-\beta}>x)Pr(1+(T/\alpha)^{-\beta}>y),\ x,y>1
\end{equation}
for $GLLD_1$ and the following one for $GLLD_2$,
\begin{equation}
Pr(1+(T/\alpha)^{\beta}>xy)=Pr(1+(T/\alpha)^{\beta}>x)Pr(1+(T/\alpha)^{\beta}>y),\ x,y>1.
\end{equation}

He also gives the Weibull's ($F(t)=1-e^{-(t/\alpha)^\beta}$) property as
\begin{equation}
Pr(X^\beta>a+b)=Pr(X^\beta>a)Pr(X^\beta>b),\ a,b>0.
\end{equation}

\section{Analogue Algebra}

\subsection{$q$-Analogues}
Results in Tsallis entropy are often efficiently expressed using $q$-analogues of elementary mathematical operations \cite{SuyTsuUes05,Yam02}.
The $q$-logarithm is defined for $q>0$ as 
\begin{align}
\label{eq:qlog}
\log_q(p) : = 
\begin{cases} 
  \log(p) & \text{if } q=1\\
  \frac{p^{1-q}-1}{1-q} &\text{otherwise}\end{cases}
\end{align}

We let the notation $\left( v \right)_+$ mean $v$ if $v>0$ and $0$
otherwise.  The inverse of the $q$-logarithm is 
\begin{equation*} 
  \exp_q (v) = \left(1 + ( 1-q) v \right)_+^{\frac{1}{1-q}} \, .
\end{equation*}
Using these two functions, we can define an analogue to multiplication:
\begin{align*}
x \qtimes y = \exp_q ( \log_q(x) + \log_q(y) )=(x^{1-q}+y^{1-q}-1)^{\frac{1}{1-q}}
\end{align*}
if $x^{1-q}+y^{1-q}-1>0$ and otherwise it is $0$. It is associative,
commutative and has $1$ as its neutral element.
Under this definition of $\qtimes$, we have the identities:
\begin{align*}
\exp_q( x + y) &= \exp_q(x) \qtimes \exp_q( y )\\
\log_q(x \qtimes y) &= \log_q(x) + \log_q(y) \text{ (whenever the left-hand-side is defined).}
\end{align*}

\subsection{Analogues Based on Link Functions}
The deformed multiplication above is a special case of defining an algebraic structure using a link functions. The idea is that if we have a set $\Omega$ and an invertible function $h$ from $\Omega$ to a set with an algebraic structure of some kind, we can pull it back using $h$ and define a corresponding structure on $\Omega$.

Suppose that we have a commutative semiring $R$, i.e. a set on which a multiplication and an addition have been defined such that the addition is commutative, associative and has an identity, the multiplication is commutative, associative, has an identity element and distributes over the addition. It is also usually part of the assumption that $0\cdot r=0$ for all $r\in R$.  

If $h:\Omega\to R$ is bijective, then we can define addition and multiplication on $\Omega$ by letting
\begin{equation}
x\htimes y=h^{-1}(h(x)h(y))
\end{equation}
and
\begin{equation}
x\hplus y=h^{-1}(h(x)+h(y)).
\end{equation}
The additive and multiplicative identities are $h^{-1}(0)$ and $h^{-1}(1)$.
We could also write that $h(x\htimes y)=h(x)h(y)$ and $h(x\hplus y)=h(x)+h(y)$.
The $q$-analogues in the previous section are defined by using the link function $h_q(x)=e^{\log_q(x)}$ to pull back the usual multiplication. If we want an addition such that $\exp_q(x\qtildeplus y)=\exp(x)\exp(y)$, then we would use
the function $\tilde{h}_q(x)=\log(\exp_q(x))$ to define $x\qtildeplus y$. The result is
\begin{equation}
x\qtildeplus y=x+y+(1-q)xy.
\end{equation}

We focus on pulling back one operation using a link function, e.g. $\qtildeplus$ using $h_q$ or we will use another link function, $\exp(\log_q(\cdot))$ to define $\qtimes$. We are really just defining semigroups for which a given link function becomes an extensive scale. We are either interested in adding data or multiplying probabilities.
 
\subsection{Memoryless Properties Based on Link Functions}
The characterizing memoryless property of a probability distribution
can be expressed by providing a function $h$, namely $h(t)=\log((F(t))^{-1})$ or alternatively the same formula but with $\bar{F}=1-F$.
We can then express this as
\begin{equation}
Pr(h(X)>a+b)=Pr(h(X)>a)Pr(h(X)>b)
\end{equation}
or
\begin{equation}
Pr(X>h^{-1}(a+b))=Pr(X>h^{-1}(a))Pr(X>h^{-1}(b)).
\end{equation}
Setting $x_0=h^{-1}(a)$ and $x_1=h^{-1}(b)$ the latter equation becomes
\begin{equation}
Pr(X>x_0\hplus x_1)=Pr(X>x_0)Pr(X>x_1).
\end{equation}

If $X$ satisfies a memoryless property based on a link function $h$, we can conclude that for some $\lambda>0$ and all $x>0$
\begin{equation}\label{exph}
Pr(h(X)>x)=e^{-\frac{x}{\lambda}}.
\end{equation}
For the Weibull distribution we have the simple function $h(x)=x^\beta$. For $GLLD_1$ and $GLLD_2$, we will consider special cases resulting from letting $\beta=1$. Letting $\frac{1}{m}=n=\alpha$ does not, as previously mentioned, result in any loss of generality.  Then we have the link functions $h_1(x)=\log({1+\frac{\alpha}{x}})^{1/\alpha}$ and $h_2(x)=\log({1+\frac{x}{\alpha}})^\alpha$. Note that $h_1(x)$ tends pointwise to $\frac{1}{x}$ when $\alpha\to 0+$ and $h_2(x)$ tends pointwise to $x$ when $\alpha\to\infty$. Ghitany preferred to express the $GLLD_1$ and $GLLD_2$ properties with $n=m=1$.

\section{Tsallis entropy}
\subsection{Deriving Tsallis Entropy}
We know that to define Tsallis entropy we want the function $V$ to satisfy
\begin{equation}
V(p_1p_2)=V(p_1)\qtildeplus V(p_2).
\end{equation}
Therefore, we would like to find a governing distribution that results in the link function that define $\qtildeplus$. We do that by using $h_2$ in the previous section that is the link function that we got from $GLLD_2$ with $\beta=1$.  

Letting $q=1-\frac{1}{\alpha}$ for $\alpha>0$ implies that $\alpha=\frac{1}{1-q}$ and 
\begin{equation}
h_2(x)=\log((1+(1-q)x)^{\frac{1}{1-q}})=\log(\exp_{q}(x))
\end{equation}
which is the desired link function. This scheme defines a governing distribution, which leads to this link function for every $q<1$. $q=1$ is simply the classical case with identity link function, which corresponds to having an exponential governing distribution.

For $GLLD_2$ with $\beta=1$, $n=\alpha>0$ and $q=1-\frac{1}{\alpha}$
\begin{equation}
Pr(X>x)=\frac{1}{(1+\frac{x}{\alpha})^\alpha}=\frac{1}{\exp_q(x)}=\exp_{2-q}(-x) .
\end{equation}
If $q>1$, we can define a probability distribution on $(0,\frac{1}{q-1})$ by
using the same formula
\begin{equation}\label{expqdist}
Pr(X>x)=\frac{1}{\exp_q(x)}
\end{equation}
which immediately results in the desired link function $\log(\exp_q)$ for $q>1$.
Thus, for all $q$, this link function comes from a distribution defined by \ref{expqdist}, however, for $q>1$ it is defined on a finite interval, which depends on  $q$.

If we base our information measure $V$ on \ref{expqdist} where
$x=\log_q\frac{1}{Pr(X>x)}$, the result is
\begin{equation}
V(p)=\log_q(\frac{1}{p})=-\log_{2-q}(p)
\end{equation}
To define the corresponding entropy, we need to calculate the expectation of $V$.
Suppose that we have a random variable with $M$ different outcomes that we are trying to model. If we have a distribution $\mu$ that assigns probability $p_i$ to outcome $i$, then its entropy is
\begin{equation}
\sum_{i=1}^M p_iV(p_i)=\sum_{i=1}^M p_i\frac{(1/p)^{1-q}-1}{1-q}=\frac{\sum_{i=1}^Mp_i^q-1}{1-q}=S_q(\mu)
\end{equation}
where we have used that $\sum_{i=1}^M p_i=1$.

\subsection{Deformed Factorization}
In this section, we consider a different way of expressing generalized memoryless properties. Equation \ref{exph} can be rewritten on a form where the link function appears on the right hand side. Suppose that
\begin{equation}
Pr(X>x)=e^{-h(x)}=\bar{F}(x).
\end{equation}
We will define a multiplication $\otimes$ such that for all $x,y>0$
\begin{equation}\label{Fmult}
\bar{F}(x)\otimes\bar{F}(y)=\bar{F}(x+y)
\end{equation}
which, with $x=\log(a)$ and $y=\log(b)$, is equivalent to
\begin{equation}
\bar{F}(\log{a})\otimes\bar{F}(\log{b})=\bar{F}(\log{ab}).
\end{equation}
Thus, we define a multiplication using the function inverse of $\bar{F}(\log(\cdot))$ as link function.
The $q$-multiplication $\qtimes$, defined in the section on $q$-analogues, is defined such that \ref{Fmult} is true when $\bar{F}=\frac{1}{\exp_q}$. Thus, the memoryless property of this special case of the generalized log-logistic can be written as
\begin{equation}
Pr(X>a+b)=Pr(X>a)\qtimes Pr(X>b).
\end{equation} 
This is an alternative to the deformation of the inside addition, which takes place in the characterization
\begin{equation}
Pr(X>x_0\qtildeplus x_1)=Pr(X>x_0)Pr(X>x_1).
\end{equation}
$\qtimes$-factorization has been used \cite{Yam02,UmaTsaSte06} to formulate a Central Limit Theorem for non-extensive statistical mechanics and a q-Hammersley-Clifford Theorem \cite{SeaSun07}.

\subsection{Generalized Statistical Theorems}
Many theorems like e.g. the Central Limit Theorem \cite{Yam02,UmaTsaSte06} have been generalized to Tsallis Statistics. This possibility is not surprising. Given the perspective of this article, we can apply the classical theorem to $h(X)$ and pull back assumptions and conclusions. 

\section{General Results}
To avoid some unnecessary technical issues we will in this section assume that all governing cumulative distribution functions are strictly increasing on the set for which its value is in the open interval $(0,1)$.
Our reasoning in previous sections has established the following theorems:
\begin{thm}\label{pullbackthm}
Suppose that $h:\Omega\to\mathbb{R}_+$ is bijective. If we define an order
on $\Omega$ by letting $x<y$ for $x,y\in\Omega$ iff $h(x)<h(y)$ in $\mathbb{R}_+$ and a group operation $\times$ on $\Omega$ by letting
$x\times_h y=h^{-1}(h(x)+h(y))$, then the resulting structure $\Omega_h$ is an ordered semigroup and
\begin{equation}
h(x\times_h y)=h(x)+h(y),
\end{equation}
or in other words, $h$ is an extensive measurement scale for $\Omega_h$.
\end{thm}

The next theorem is our way of formulating one direction of Ghitany's main theorem.
\begin{thm}[Ghitany]
If $F$ is a cumulative distribution function on $(-\infty,\infty)$, then there is a function $h$ from $\Omega=\{x\in\mathbb{R}\ |\ F(x)\in (0,1)\}$ to $\mathbb{R}_+$ such that $Pr(h(X)>x+y)=Pr(h(X)>x)Pr(h(X)>y)$ for $x,y>0$, which implies that there is $\lambda>0$ such that $Pr(h(X)>x)=e^{-\lambda x}$ and
$Pr(X>y)=e^{-\lambda h(y)}$ for $y\in\Omega$.
\end{thm}

Suppose we have an interval $\Omega$ in $\mathbb{R}$, a governing distribution $F$ with survival function $\bar{F}$ and a link function $h:\Omega\to\mathbb{R}_+$ such that $Pr(h(X)>x+y)=Pr(h(X)>x)Pr(h(X)>y),\ x,y>0$. The existence of $h$ is guaranteed by Ghitany's Theorem. Theorem \ref{pullbackthm} provides us with an ordered semigroup structure on $\Omega$, which we denote by $\Omega_h$. Denoting the group operation by $\oplus_h$, we can express the memoryless property of $F$ by
$$Pr(X>x\oplus_h y)=Pr(X>x)Pr(X>y).$$

\begin{thm}\label{entthm}
Suppose that $V:(0,1]\to\Omega$ is defined by $V(\cdot)=\bar{F}^{-1}(\cdot)$ with definitions as above. Then $V(p_1p_2)=V(p_1)\oplus_h V(p_2)$ and furthermore, if $(A,\mu)$ is a probability space and we define an entropy by letting $S(\mu)=E_\mu(V)$ it is extensive with respect to $\oplus_h$. 
\end{thm}  

Theorem \ref{entthm} is saying that a governing distribution provides a link function, an ordered semigroup structure with respect to which $h$ is an extensive measurement scale and an entropy, which is extensive with respect to $\Omega_h$.

\begin{thm}
We have also seen that if we let $\tilde{h}(\cdot)=\bar{F}(\log(\cdot))$ and let \\$p_1\otimes_{\tilde{h}^{-1}}p_2=\tilde{h}(\tilde{h}^{-1}(p_1)\tilde{h}^{-1}(p_2))$, we can express the memoryless property of $F$ by 
\begin{equation}
Pr(X>x+y)=Pr(X>x)\otimes_{\tilde{h}^{-1}}Pr(X>y).
\end{equation}
\end{thm}

\section{A Rainfall Example}
We have in this article used rainfall modeling as an example. Hydrological modeling is a whole science in itself and we will here just give a concrete example with rainfall data from Spey(at Kinrara) in Scotland, previously studied in many papers including \cite{AhmSinWer88} where log-logistic models are applied. The data is an annual maximum series in $m^3s^{-1}$ for the years 1952-1982. The series is: 

$89.8,\ 109.1,\ 202.2,\ 146.3,\ 212.3\ 116.7\ 109.1,\ 80.7,$ $127.4,\ 138.8,\ 283.5,\ 85.6$ $105.5,\ 118.0\ 387.8,\ 80.7,\ 165.7$ $111.6,\ 134.4,\ 131.5,\ 102.0,\ 242.5,\ 214.8,\ 144.6,$ $114.2,$ $98.3,\ 102.8,\ 104.3,\ 196.2,\ 143.7$.

To model such data it is common to combine data from hydrologicaly homogeneous zones to find enough data to estimate parameters from. We will not try to estimate anything but just give an example of what a Tsallis scale transformation is doing. If we would let $X$ be the difference between the annual maximum and $60$, which is approximately half the median for the data, we would consider the functions $\frac{1}{1-q}\log(1+(1-q)X)$. Since $e^{(1-q)x}\approx 1+(1-q)x$ if $(1-q)x$ is close to zero, we are not changing the scale by much as long as we are close to zero, while larger values are rescaled more substantially.
If we apply this transformation with $q=\frac{1}{2}$, the resulting sequence has as its smallest elements $2.1,\ 2.1,\ 2.3,\ 2.4$ and the largest $4.4,\ 4.1,\ 3.9,\ 3.8$. We have arrived at a more tempered scale. We do not, however, claim that it truly is an extensive measurement scale for the studied phenomenon.

\section{Remarks}

\subsection{Events at Different Levels}
Sometimes the word scale is used in a different sense then it has been used in this article, e.g. subatomic, atomic, molecular, cellular, organism, sociological, ...,  astronomical scale. An interesting property of the Tsallis entropy scale, which is based on the distribution with $Pr(X>x)=\frac{1}{(1+\frac{x}{\alpha})^\alpha}$, is that if $\alpha=n$ is a natural number and
$Pr(X_i>t)=\frac{1}{1+t}$, then
\begin{equation}\label{subev}
Pr(X>x)=\prod_{i=1}^n Pr(X_i>\frac{x}{n}).
\end{equation}
If $\{X_i\}$ are independent and we let $X=n\min_i X_i$, then $X$ satisfies \ref{subev}. This situation could arise if $X_i$ are components in a chain, which produces $n$ times more units of something, e.g. energy or money, than a single component but which also has the property that the entire chain stops if any component breaks. A similar situation would occur if we need equal amounts of $n$ different substances to make a certain product. The residues are not observable if we only see the resulting product. This is typical when we observe a complex multi-level system at a specific level. If we would look at this from a superstatistics point of view where we work with Gamma distributions, we note that a Gamma distribution with integer shape parameter  $k$ is the distribution of a sum of $k$ exponentially distributed random variables. It is also true that if we add Gamma distributed random variables with the same scale parameter, the result is Gamma distributed with the same scale but with a shape that is the sum of the shapes of the terms. These properties are suitable if we want to model observations which are really combinations of events at a smaller unobserved level.

\subsection{Radial Basis Functions}
A Radial Basis Function is a function $h:\mathbb{R}_+\to\mathbb{R}_+$. The idea is to transform  the euclidean distance by
considering $h(|x-y|)$ instead of $|x-y|$, i.e. it is about changing the scale. It is often used in interpolation theory when we want to approximate a function $f:\mathbb{R}^d\to \mathbb{R}$ from a finite number of (possibly approximate) function values $f(x_i)$. The rescaling of the norm can be expressed as using a feature map into a Reproducing Kernel Hilbert Space (RKHS) with kernel $k(x,y)=h(|x-y|)$. A common choice is the Gaussian $h(r)=e^{-r^2/\sigma}$. It focuses the influence of a function value much more towards the immediate surroundings of the point, while it almost extinguishes its long range influence. RKHS are used in many application areas including Chemical Physics where they are used to construct multidimensional molecular potential energy surfaces \cite{HoRab96}.

\section{Summary}
We have defined an extensive measurement scale for a random variable $X$ as being a function $h$ which makes the distribution of $h(X)$ memoryless. An alternative to applying $h$ to $x$ is to deform the elementary mathematical operations that we use to analyze $X$. This includes deforming the entropy. By choosing different governing distributions, a general class of entropies and information measures arise. We show that Tsallis entropy can be derived in this way from a class of generalized log-logistic distributions.



\bibliographystyle{elsart-num}   

\bibliography{qbib}
\IfFileExists{\jobname.bbl}{}
 {\typeout{}
  \typeout{******************************************}
  \typeout{** Please run "bibtex \jobname" to optain}
  \typeout{** the bibliography and then re-run LaTeX}
  \typeout{** twice to fix the references!}
  \typeout{******************************************}
  \typeout{}
 }


\end{document}